\documentclass[a4paper,12pt]{article}
\usepackage{epsfig}
\usepackage[titletoc,title]{appendix}
\usepackage{epstopdf}\usepackage[sumlimits,intlimits,namelimits]{amsmath}
\usepackage{dsfont}
\usepackage{slashed}
\usepackage{color}

\usepackage[english]{babel}

\usepackage{graphicx}

\usepackage{geometry}
\makeatletter
\def\fmslash{\@ifnextchar[{\fmsl@sh}{\fmsl@sh[0mu]}}
\def\fmsl@sh[#1]#2{%
  \mathchoice
    {\@fmsl@sh\displaystyle{#1}{#2}}%
    {\@fmsl@sh\textstyle{#1}{#2}}%
    {\@fmsl@sh\scriptstyle{#1}{#2}}%
    {\@fmsl@sh\scriptscriptstyle{#1}{#2}}}
\def\@fmsl@sh#1#2#3{\m@th\ooalign{$\hfil#1\mkern#2/\hfil$\crcr$#1#3$}}
\makeatother

\newcommand{\ep}{\epsilon}

\newcommand{\ice}[1]{\relax}

\numberwithin{equation}{section}
\begin{document}
\begin{titlepage}
\begin{flushright}
SI-HEP-2017-13 \\
QFET-2017-10
\end{flushright}

\begin{center}
{\Large\bf
Towards a Next-to-Next-to-Leading Order analysis \\[2mm]
of matching in \boldmath $B^0$--$\bar{B}^0$ \unboldmath mixing
 }
\end{center}

\begin{center}
{\sc Andrey G. Grozin} \\[2mm]
{\sf  Budker Institute of Nuclear Physics SB RAS, Novosibirsk 630090,
  Russia}
and \\
{\sf Novosibirsk State University, Novosibirsk, 630090, Russia}
 \\[3mm]
{\sc Thomas Mannel } and {\sc Alexei A. Pivovarov} \\[0.1cm]
{\sf Theoretische Elementarteilchenphysik, Naturwiss.--techn. Fakult\"at, \\
Universit\"at Siegen, 57068 Siegen, Germany}
\end{center}

\begin{abstract}
\noindent
We compute perturbative corrections to matching coefficients of
QCD to heavy quark effective theory (HQET)
for the matrix element of the $\Delta B=2$ operator
that determines the mass difference in  $B^0-\bar{B}^0$
system of states.
This involves the technical point of
choosing the operator basis in  HQET, separating physical
operators from evanescent ones. We obtain an analytical
result for some of the two-loop  corrections
at order $\alpha_s^2$.

\end{abstract}


\end{titlepage}

\pagenumbering{arabic}

\section{Introduction}
\label{Sect:Intro}
Theoretical predictions
within the Standard Model (SM) of particle physics
for the oscillation frequency
and the lifetime differences in the systems of neutral bottom mesons
have become much more accurate
during the last decade~\cite{Lenz:2006hd,Lenz,Nierste}.
This progress is pretty much related to the numerical computation
of the relevant hadronic matrix elements using
lattice simulations of QCD
(e.g.~\cite{Aoki:2016frl,Aoki:2014nga}).
$B^0$-$\bar{B}^0$ oscillations are
flavor-changing neutral current processes and hence
are sensitive to possible effects beyond the SM,
so our ability to perform precise calculations
is of prime
importance~\cite{Carrasco:2013zta,Bazavov:2016nty}.
In fact, the high precision of experimental data
provides us with good opportunities
for searches of physics beyond the SM~\cite{Dowdall:2014qka},
in particular in
the mixing of states
in the systems of neutral flavored mesons.
The mixing
in the neutral kaon and the charmed-meson
systems is strongly
long-distance sensitive and thus depends
on nonperturbative QCD effects. This is in contrast to
the oscillation frequency in
the neutral
$B$ mesons which
is dominated by the top-quark contribution and hence
it is short distance dominated even in the SM, thereby making
a clean
SM prediction possible.

Highly precise predictions for observables in the $B$-meson
system are based on the method of effective field theory (EFT),
which allows us to separate vastly different mass scales and to
perform an expansion in the small ratios of these
scales~\cite{Inami:1980fz,Buras:1998raa,Grozin:2013hra}.
This involves, on the one
hand, the use of EFT for weak interactions as an
expansion in $m_b / M_{W/Z}$
and $m_b / m_t$, and,
on the other hand, the use of heavy quark effective theory (HQET) as an expansion in
$\Lambda_{\rm QCD} /m_b$. However,
as soon as the scales become so low that a perturbative evaluation
is impossible, nonperturbative input becomes necessary in form of
hadronic matrix elements. To this end, observables in the $B$-meson
system can be
generically written as perturbatively computable Wilson coefficients
and low-energy hadronic
matrix elements~\cite{Buras:1990fn,BBL:96}.
While the Wilson coefficients
have rather accurate
theoretical predictions within perturbation
theory~\cite{Beneke:1998sy}
the recent progress in lattice produces
the comparable accuracy of computation
for matrix elements.

However, the data to be expected in
the near future will require to further
improve the accuracy. This eventually
requires even more precise values of Wilson coefficients
as well as improved
lattice calculations.  Clearly, one of the most crucial
points of such an improvement
are the perturbative corrections at the lowest scales
involved in the analysis.
As the first step of
this program one has to compute the
next-to-next-to leading
order (NNLO) corrections
to the matching of QCD to heavy quark effective theory (HQET)
that is applicable at the scales of order few $\Lambda_{\text{QCD}}$.
In the present paper we discuss the choice of the operator
basis within dimensional regularization
and give explicitly some of the NNLO contributions.

The paper is organized as follows.
We give a
short necessary description of the procedure of computation of
the effective Hamiltonian at the bottom quark mass scale $m_b$
and definition of the
operator basis for four-quark operators
with introduction of evanescent operators in QCD.
Then we define the operator basis in HQET and discuss the
correspondence
of the two
bases. With two  bases at hand we consider a simple computation of
matching coefficients with
fermion bubbles to set the stage for the full
computation at NNLO. In Sect.~\ref{Sect:SR} we consider the
contribution of evanescent operators to the correlator for the
sum rules analysis in HQET.
In Sect.~\ref{Sect:Flavor} we discuss the bag parameter for $B_s$
meson
within unitary symmetry with expansion in the strange quark mass.
Sect.~\ref{Sect:Summary} contains our conclusions.

\section{Wilson coefficients and operator
basis}
\label{Sect:basis}
In the SM the amplitude for $B^0$-$\bar{B}^0$ transitions
vanishes at tree level. The leading contribution are the box diagrams
appearing at the one-loop level. The dominant part is induced by the loop
diagram involving the top quark and the $W$ boson. On the other hand,
the relevant scale for this process is much lower, of the order of the
$b$-quark mass. Switching to an EFT for weak interactions allows us
to integrate out the top quark and the $W$ boson, leaving us with
local four-quark operators.

The most important corrections to the leading term
are the contributions of
strong interactions. They can be computed within QCD perturbatively
as the relevant scale is of the order
of the meson mass $m_B$ and is much larger
than QCD infrared scale $\Lambda_{\text{QCD}}$.
Finally, the matrix element of the four-quark
operator still contains the scale $m_b$,
which can be treated in QCD perturbation theory. By switching
from QCD to HQET
we may extract also these (perturbative) contributions,  leaving
us with matrix elements
of static HQET fields.

\subsection{QCD: From $M_W$ to $m_b$}

The scales between $M_W$, $m_t$ and $m_b$ are in a perturbative
domain for strong interactions.
After integrating the large scales $M_W$, $m_t$ out perturbatively,
the process of  $B^0$--$\bar{B}^0$ is described by
an effective Hamiltonian of the form
\begin{equation}
{\cal H}=C(M_W, m_t,\mu, \alpha_s(\mu))Q(\mu)
\label{eq:HeffWeak}
\end{equation}
with
\begin{eqnarray}
Q(\mu)=(\bar q_L \gamma_\alpha b)\, (\bar q_L \gamma_\alpha b)(\mu)
\label{eq:Q}
\end{eqnarray}
being a local operator with $\Delta B=2$
renormalized at the scale $\mu$.
The amplitude for $B^0$-$\bar{B}^0$  mixing is obtained by
computing a matrix
element of  (\ref{eq:HeffWeak}) between $B^0$ and $\bar{B}^0$
states.  Choosing the scale $\mu\sim m_b$
as the renormalization scale will avoid the appearance
of the large scales $M_W$ and $m_t$ in the matrix elements; these dependencies
reappear in
the Wilson coefficient
$C(m_W, m_t,\mu,\alpha_s(\mu))$ that
accounts for the effects of the evolution between $m_t$, $M_W$
and
$\mu\sim m_b$, or just resums those large
logarithms of the scales ratio, $\ln(M_W/m_b)$.
It is a standard tool to use the
renormalization group (RG) to evolve the scale
$\mu$ from $m_t$, $M_W$ to  $m_b$, thereby resumming  large
logarithms of the scales ratio, $\ln(M_W/m_b)$. The RG technique
requires the calculation of the $\beta$
function governing the running of $\alpha_s$
and the anomalous dimensions $\gamma$, which can be done in QCD
perturbation theory. Using the one-loop results for $\beta$ and $\gamma$ in
combination with tree-level matching at $\mu = M_W$ yields the leading order result, while
the input of $\beta$ and $\gamma$ at two loops
together with the one-loop matching
yields the result at next-to-leading order (NLO).

At present the coefficient $C(M_W, m_t,\mu,\alpha_s(\mu))$
in (\ref{eq:HeffWeak})
is known at the NLO of the expansion in the strong
coupling constant that gives the accuracy of a few
percent~\cite{Buras:1990fn,BBL:96,Beneke:1998sy}.
The RG function $\beta$ and $\gamma$ are related to divergences in Feynman diagrams,
which need to be regularized.
Technically,
the best way of regularization in perturbative multiloop
calculations is
dimensional regularization (DimReg)
where the Feynman diagrams are computed in $d$
dimensions.
However, the mixing process in the SM
contains left-handed quark fields and DimReg has no
unique or even simple satisfactory way of treating the Dirac matrix
$\gamma_5$. This problem is vital for supersymmetric theories
and some solutions
have been suggested. Dimensional reduction allows for the
computation of corrections to $\Delta F=1$
Hamiltonian~\cite{Altarelli:1980fi}.
There are also techniques by
't~Hooft and Veltman (HV)~\cite{tHooft:1972tcz}
and a naive dimensional regularization
prescription (NDR)~\cite{Chanowitz:1979zu}.
For four-quark operators one can use techniques of
refs.~\cite{Buras:1989xd,Dugan:1990df,Herrlich:1994kh}
that are
a practical way out through
an extension of the operator basis by using so-called
evanescent operators.
Note that the two-loop
anomalous dimensions of baryon operators,
or three-quark operators, have been computed earlier within
a similar approach~\cite{Pivovarov:1988gt,Pivovarov:1991nk}.
A clear
presentation of the techniques
is given in~\cite{Chetyrkin:1997gb}.

In an extended operator basis
with evanescent operators $E_i(\mu)$
the effective Hamiltonian is written as
\begin{eqnarray}\label{eq:ext-ham}
{\cal H}=C(M_W, m_t,\mu, \alpha_s(\mu))Q(\mu)+\sum_i C_{E_i}
E_i(\mu)\, .
\end{eqnarray}
The number of evanescent operators and their structures
depend on the order of perturbation theory to which the Hamiltonian is
computed.
The appearance of evanescent operators is an artifact of using
dimensional regularization and these operators play an auxiliary
role.
The generic feature of evanescent operators
is that they are, in a sense, equivalent to zero after going to four-dimensional space-time, $d\to 4$.
More precisely, these operators are defined such that
their Green functions with
fundamental quark-gluon fields
vanish in perturbation theory for $d=4$
\begin{eqnarray}\label{eq:evan-vanish}
\langle E_i(\mu)\rangle=0\, .
\end{eqnarray}
Here the brackets mean to compute
any Green function of the operators $E_i(\mu)$ and renormalized
quark and gluon fields within perturbation theory (PT)
in an infrared safe kinematical point.
In practice, one computes an
amputated connected Green's function of the operators
$ E_i(\mu) $  with four fermionic fields
that is sufficient for extracting the UV properties of the
operator basis in eq.~(\ref{eq:ext-ham}).
However, the coefficient $C(M_W, m_t,\mu, \alpha_s(\mu))$
of the physical operator $Q(\mu) $
and also the very definition of $Q(\mu) $ itself
depends on
the choice of the whole basis $\{Q(\mu),E_i(\mu)\}$,
and, therefore, the concrete set of evanescent operators $E_i(\mu)$.
Physical operator(s) $Q(\mu)$ mix with the
evanescent operators $E_i(\mu)$ under renormalization.
For finite renormalization it means that with the change of
the renormalization parameter one has
\begin{eqnarray}
\left(
\begin{array}{c}
 Q (\mu^\prime)   \\
 E (\mu^\prime)
\end{array}\right)
=
\left(
\begin{array}{cc}
 Z_Q (\mu^\prime ,\mu ) & Z_{QE} (\mu^\prime ,\mu ) \\
 Z_{EQ} (\mu^\prime , \mu ) &  Z_{EE} (\mu^\prime , \mu )
\end{array}\right)
\left(\begin{array}{c}
Q(\mu) \\
E(\mu)
\end{array}\right)
\end{eqnarray}
and for the physical sector
\begin{eqnarray}
 Q (\mu^\prime)  =
Z_Q (\mu^\prime ,\mu )  Q(\mu)+Z_{QE} (\mu^\prime ,\mu )  E(\mu)\, .
\end{eqnarray}
Here both $Q$ and $E$ can be columns of operators and any $Z$
is a matrix.
In terms of bare operators $(\mathrm{Q},\mathrm{E})$
the renormalization reads
\begin{equation}
\left(\begin{array}{c}\mathrm{Q}\\\mathrm{E}\end{array}\right)
= Z
\left(\begin{array}{c}Q(\mu)\\E(\mu)\end{array}\right)\,.
\end{equation}
At one loop the renormalization constant $Z$ has the form
\begin{equation}
Z = 1 + \frac{\alpha_s}{4\pi\varepsilon}
\left(\begin{array}{cc}
z_{QQ} & z_{QE} \\
\epsilon z_{EQ} & z_{EE}
\end{array}\right)\,,
\end{equation}
where $z_{EQ}$ is obtained from the requirement
that corresponding Green's functions
of the renormalized evanescent operator $E(\mu)$, as
in eq.~(\ref{eq:evan-vanish}), vanish.

In the following we shall use, as a short-hand notation, a
symbol of direct product $\Gamma\otimes \Gamma$
for the Dirac structure of four-quark operators.
Thus, one has
\begin{eqnarray}
\Gamma\otimes \Gamma\equiv \bar q_L \Gamma b \bar q_L \Gamma b\, .
\end{eqnarray}
The color structure of the operator
can be also indicated in this manner.
It can be either $1\otimes 1$ or $t^a\otimes t^a$, or some mixture of
both (see below).
If not explicitly said, the short-hand notation tacitly implies
that a
relation between operators
is independent of color arrangement of quark fields.

The evanescent operators of dimension six with four-quark structure
naturally emerge for DimReg
treatment of flavor mixing processes
since the Dirac algebra of gamma matrices
becomes formally
infinite dimensional in $d=4-2\ep$ dimensional space-time for
arbitrary $\ep$. Therefore,
the basis in Dirac algebra contains
the products of gamma matrices with any number of those.
Higher products of gamma matrices are not reducible to lower ones
anymore contrary to four-dimensional case.
The first example emerges already with the product of three gamma
matrices. In four dimensions, the relation
\[
\gamma_\mu\gamma_\nu\gamma_\alpha=g_{\mu\nu}\gamma_\alpha+
g_{\nu\alpha}\gamma_\mu - g_{\mu\alpha}\gamma_\nu
+i\ep_{\mu\nu\alpha\beta}\gamma_\beta\gamma_5
\]
allows to reduce the number of gamma matrices in the product.
This is not possible in $d$ dimensions anymore.
However, for four-quark operators with left quarks that are
eigenstates of chirality one can simply parametrize the
difference between a higher product of  gamma matrices and its
four-dimensional limit with a new object that is called an evanescent
operator.
Thus, while in four dimensions there exists a reduction
of the form
\begin{eqnarray}\label{eq:reduction4}
\gamma_\mu\gamma_\nu\gamma_\alpha\otimes
\gamma_\mu\gamma_\nu\gamma_\alpha
\to 16\gamma_\alpha\otimes \gamma_\alpha \, ,
\end{eqnarray}
in $d$ dimensions it becomes a relation
\[
\gamma_\mu\gamma_\nu\gamma_\alpha\otimes
\gamma_\mu\gamma_\nu\gamma_\alpha
\to 16\gamma_\alpha\otimes \gamma_\alpha
+E^{\rm evan}
\]
that defines an evanescent operator $E^{\rm evan}$.
Note also that
the renormalization of four-quark operators would require
an introduction of evanescent operators even in pure
vectorlike theories if DimReg is used.

The choice of the evanescent operator $E^{\rm evan}$ is not unique
(see, e.g.~\cite{Herrlich:1994kh}).
A concrete recipe of treating evanescent operators in
QCD for flavor changing processes within naive
dimensional regularization
has been formulated in ref.~\cite{Buras:1989xd} where Wilson
coefficients at NLO have been analysed in detail.
At a computational level the particular recipe of
ref.~\cite{Buras:1989xd} reduces to
a substitution
\[
\gamma_\mu\gamma_\nu\gamma_\alpha\otimes
\gamma_\mu\gamma_\nu\gamma_\alpha
\to (16-4\epsilon)\gamma_\alpha\otimes \gamma_\alpha
+E^{QCD}\]
with $d=4-2\epsilon$.
Thus, the evanescent operator in QCD is defined in the original
publication~\cite{Buras:1989xd} as
\begin{eqnarray}\label{eq:QCD-evan}
E^{QCD}=\gamma_\mu\gamma_\nu\gamma_\alpha\otimes
\gamma_\mu\gamma_\nu\gamma_\alpha
-(16-4\epsilon)\gamma_\alpha\otimes \gamma_\alpha \, .
\end{eqnarray}
An infrared safe Green's function of (renormalized)
$E^{QCD}(\mu)$ being computed
over the quark states vanishes in $d=4$
in perturbation theory up to NLO.
In higher orders of PT new evanescent operators will appear and the
coefficient of the physical structures will have higher orders of
$\ep$ expansion.
The choice  of $E^{QCD}$ is not unique.
The actual choice of the basis in ref.~\cite{Buras:1989xd}
is a nonminimal choice of the basis in a sense that there is an
addition of explicit order $\ep$ of a physical operator to the
reduction relation~(\ref{eq:reduction4}).
This choice is motivated by the requirement of
validity of Fierz transformation in the physical
sector and
discussed in detail in~\cite{Buras:1989xd,Herrlich:1994kh}.

Note that the one-loop counterterm to the operator $Q$
has the form~\cite{KOPP:03}
\begin{eqnarray}
(\bar q_L t^a\Gamma^{[3]}b)( \bar q_L t^a\Gamma^{[3]}b)
\end{eqnarray}
where $\Gamma^{[3]}$ is a totally
antisymmetric product of three Dirac gamma
matrices, and $t^a$ are $SU(3)$ color generators of QCD.
The totally antisymmetric products of $n$ Dirac gamma matrices
$\Gamma^{[n]}$
form also a convenient basis in the Dirac algebra
in $d$-dimensions.
In terms of this basis the above evanescent operator in QCD
reads
\begin{eqnarray}\label{eq:antybasis}
E^{QCD}&=&\gamma_\mu\gamma_\nu\gamma_\alpha\otimes
\gamma_\mu\gamma_\nu\gamma_\alpha
+ (-16+4\epsilon)\gamma_\alpha\otimes \gamma_\alpha
\nonumber \\
&=&
\Gamma^{[3]}\otimes\Gamma^{[3]}
+(3d-2)\Gamma^{[1]}\otimes\Gamma^{[1]}
+(-16+4\epsilon)\Gamma^{[1]}\otimes\Gamma^{[1]}
+ O(\epsilon^2)\nonumber \\
&=&\Gamma^{[3]}\otimes\Gamma^{[3]}
+(-6-2\epsilon)\Gamma^{[1]}\otimes\Gamma^{[1]}+O(\epsilon^2)
\end{eqnarray}
with $\Gamma^{[1]}\otimes\Gamma^{[1]}\equiv
\gamma_\nu\otimes\gamma_\nu$.
The anomalous dimension of the operator $Q(\mu)$ and the Wilson
coefficient
$C(\mu)$
at NLO in QCD
are given in the literature
mainly for this particular choice of the basis~(\ref{eq:QCD-evan}).

The renormalized operator $Q(\mu)$
(in contrast to the bare one $\mathrm{Q}$)
depends on the choice of the evanescent
one~\cite{Herrlich:1994kh,Chetyrkin:1997gb,Herrlich:1996vf,Gorbahn:2004my}.
If we choose $\mathrm{E}' = \mathrm{E} + a \epsilon \mathrm{Q}$
instead of $\mathrm{E}$,
we obtain a different renormalized operator $Q'(\mu)$.
At the one-loop level the relation between the two operators goes as
\begin{equation}
Q'(\mu) = \left[1 - a z_{QE} \frac{\alpha_s(\mu)}{4\pi}\right] Q(\mu)\,.
\label{PhysEva}
\end{equation}

\subsection{HQET: Below $m_b$}
\label{Sect:basisHQET}
The second step in the effective theory analysis of the $B^0$--$\bar{B}^0$
mixing is
to remove an explicit dependence on $m_b$
from the matrix element or the mixing amplitude at low energy since
this scale is still QCD perturbative, $m_b\gg \Lambda_{\text{QCD}}$.
The removal of $m_b$ scale
is achieved by using HQET~\cite{N:94,MW:00,G:04}.
One requires
matching QCD to HQET at the scales around $m_b$.
At scales $\mu$ below the $b$ quark mass the QCD operators
are expanded into a series of local HQET
operators. This is called heavy quark expansion
(HQE) and  symbolically denoted as
expansion in
$\Lambda_{\text{QCD}} / m_b$.

In particular, the HQE of the operator $Q(\mu)$ goes
\begin{equation}
Q(\mu) = 2 \sum_{i=l,s} C_i(\mu) {O}_i(\mu) +
\mathcal{O}\left(\frac{\Lambda_{\text{QCD}}}{m_b}\right)\,
\label{eq:match}
\end{equation}
where the HQET operators ${O}_{l,s}(\mu)$
are defined as
\begin{eqnarray}\label{eq:original_matching}
O_l=(\bar q_L \gamma_\mu h_+)( \bar q_L \gamma_\mu h_- ),\quad
O_s=(\bar q_L h_+)( \bar q_L h_-)\, .
\end{eqnarray}
The bare field $h_+$ annihilates the heavy quark in HQET
(moving
with the four velocity $v$),
and $h_-$ creates the heavy antiquark (again moving with
the four velocity $v$),
which is a completely separate particle in HQET framework.
Note that a single physical operator of QCD, $Q(\mu)$, is expanded over
two independent operators of HQET,  ${O}_{l,s}(\mu)$.

In general one has to define in HQET
its own set of physical and evanescent operators.
It is convenient to choose four-quark operators
in analogy
with QCD. We introduce
\begin{eqnarray}\label{eq:gen-oper-HQET}
O_n=(\bar q \gamma_\perp^{[n]} h_+)( \bar q \gamma_\perp^{[n]} h_-
),
\quad
O_n^\prime=(\bar q_i \gamma_\perp^{[n]} h_+^j)
( \bar q_j \gamma_\perp^{[n]} h_-^i ),\quad
\end{eqnarray}
and $q$ is
always a left-handed fermion $q\equiv q_L$. It is usually a light-flavored one, $d$ or $s$.
A natural choice for a basis in HQET
is an antisymmetrized product of transverse gammas,
\begin{eqnarray}
\gamma_\perp^\mu=\gamma^\mu-v^\mu\slashed{v}\, .
\end{eqnarray}
Then  $\gamma_\perp^{[n]}$ is an antisymmetrized product of transverse
gamma matrices with a rank (number of matrices)
$n$.
We use further
notation: the capital $\Gamma^{[n]}$ is full antisymmetric as in QCD,
small $\gamma^{[n]}_\perp$ is transverse.
Then
\begin{eqnarray}
\Gamma^{[n]}\otimes\Gamma^{[n]} &=&
\gamma_\perp^{[n]}\otimes\gamma_\perp^{[n]}
-n\gamma_\perp^{[n-1]}\otimes\gamma_\perp^{[n-1]}
\end{eqnarray}
in a sense of four-quark operators in HQET
of the form eq.~(\ref{eq:gen-oper-HQET}).
Up to NLO computation (one loop)
one can meet in the calculations
a product of not more than three gamma matrices only (and four in HQET).

The simplest basis of physical and evanescent operators
emerges after a direct reduction of four-quark operators
in four dimensions with the use of the Fierz rearrangement.
With $O_0,O_1$ from~(\ref{eq:gen-oper-HQET})
being chosen as a physical pair
the four-dimensional reduction of the operators of the form given in
eq.~(\ref{eq:gen-oper-HQET})
reads
\begin{eqnarray}
O_2&=&-2O_1, \quad O_3=-6O_0, \nonumber   \\
O^\prime_0&=&-\frac{1}{2}(O_0+O_1), \quad
O^\prime_1=-\frac{3}{2} O_0+\frac{1}{2}O_1, \nonumber   \\
O^\prime_2&=&3 O_0-O_1, \quad
O^\prime_3=3 O_0+3O_1 \, .
\end{eqnarray}
Thus
a set of corresponding evanescent operators reads
\begin{eqnarray}\label{eq:basis-HQET}
e_1&=&O_2+2O_1, \quad
e_2=O_3+6O_0, \nonumber   \\
e_3&=&O^\prime_0+\frac{1}{2}(O_0+O_1), \quad
e_4=O^\prime_1+\frac{3}{2} O_0-\frac{1}{2}O_1, \nonumber   \\
e_5&=&O^\prime_2-3 O_0+O_1, \quad
e_6=O^\prime_3-3 O_0-3O_1
\end{eqnarray}
and $O_0,O_1$ form a physical pair.
Other convenient choices of the bases are discussed in
Appendix~\ref{app:basis} in some detail.
Here we only mention that one possible physical basis is also
\begin{eqnarray}
O_l=\Gamma^{[1]}\otimes\Gamma^{[1]} =O_1-O_0,
\quad O_s=\Gamma^{[0]}\otimes\Gamma^{[0]}=1\otimes 1 = O_0\, .
\end{eqnarray}
This basis is such that the operator $ O_s$
appears only at NLO
in matching.
It has been used in original papers on matching QCD onto
HQET~\cite{FHH:91,B:96,CFG:96}
and is indicated in eq.~(\ref{eq:original_matching}).

The basis~(\ref{eq:basis-HQET}) in HQET
is fully appropriate for further use in HQET
but it does not match a QCD evanescent
definition of ref.~\cite{Buras:1989xd}
in a sense that the evanescent operator $E^{\rm QCD}$
does not match onto evanescent operators of an HQET
basis~(\ref{eq:basis-HQET}).
While it is not a necessary requirement such a property would be
convenient in practical computation.

Let us work out the basis which has such a property.
An expansion of antisymmetric product of gammas relevant for QCD
over the HQET basis of transverse
products
reads
\begin{eqnarray}
\gamma_\mu\gamma_\nu\gamma_\alpha\otimes
\gamma_\mu\gamma_\nu\gamma_\alpha
&=&\Gamma^{[3]}\otimes\Gamma^{[3]}
+(3d-2)\Gamma^{[1]}\otimes\Gamma^{[1]}\\
&=& \gamma_\perp^{[3]}\otimes \gamma_\perp^{[3]}
-3 \gamma_\perp^{[2]}\otimes \gamma_\perp^{[2]}+
(3d-2)( \gamma_\perp^{[1]}\otimes \gamma_\perp^{[1]}
- \gamma_\perp^{[0]}\otimes \gamma_\perp^{[0]})
\end{eqnarray}
where $ \gamma_\perp^{[0]}\otimes \gamma_\perp^{[0]}\equiv 1\otimes
1$.
Any basis is ``legal'' to use in the calculation.
In some bases, however, the
evanescents give a
contribution to a physical sector of HQET due to radiative corrections
and have to be explicitly matched.
One can choose a basis when QCD evanescent matches to
HQET evanescent. Of course, this happens
with some accuracy in $\ep$ expansion only.
In our case
one sees that it suffices to shift the definition of the highest
rank operator ($O_3$) with both color
structures in accordance with the
prescription of ref.~\cite{Buras:1989xd}.
Then  the canonical QCD evanescent operator
matches to HQET evanescent.

Finally, our working basis in HQET is $\{O_0,O_1,e_i\}$
with
\begin{eqnarray}\label{eq:HQET-good-basis}
O_2&=&-2O_1+e_1, \nonumber   \\
O_3&=&-6O_0+2\ep(O_1-O_0)+e_2+O(\ep^2), \nonumber   \\
O^\prime_0&=&-\frac{1}{2}(O_0+O_1)+e_3, \nonumber   \\
O^\prime_1&=&-\frac{3}{2} O_0+\frac{1}{2}O_1+e_4, \nonumber   \\
O^\prime_2&=&3 O_0-O_1+e_5, \nonumber   \\
O^\prime_3&=&3(O_0+O_1)+2\ep(O_1-O_0)+e_6+O(\ep^2)\ .
\end{eqnarray}
With such a choice the QCD canonical evanescent operator $E^{QCD}$
eq.~(\ref{eq:QCD-evan})
matches to pure
HQET evanescent operators with necessary accuracy
in $\ep$-expansion.

\section{Matching computation}
\label{Sect:HQET}
After fixing the choice of the basis we compute the matching
coefficients $C_i$ in eq.~(\ref{eq:match}).
The basis in terms of transverse antisymmetric
gammas allows for projecting
diagrams and completely automatic handling of the whole computation.
We use the symbolic system REDUCE for these calculations.

The computation has been done in leading logs
in~\cite{PW:88,SV:88} where LO anomalous dimension has been
found.
It happens to be equal to twice the anomalous dimension
of a bilinear current. In higher orders it is not so.
The standard result at NLO is reproduced~\cite{FHH:91,B:96,CFG:96}.
Note that the use of the minimal basis of eq.~(\ref{eq:basis-HQET})
gives a different answer for the NLO matching coefficient
and therefore requires an explicit matching of the QCD
evanescent operator $E^{QCD}$ for reproducing
the result given in the literature.
The adjusted basis of eq.~(\ref{eq:HQET-good-basis}) reproduces the
correct NLO matching coefficient automatically.
This is a clear difference with computation of anomalous dimensions.
While the presence of evanescent operators starts to
influence the computation
of anomalous dimensions at two-loop level, the matching coefficients
are sensitive to the particular basis of evanescent operators already
at one-loop level.

In this paper, as a first step of NNLO calculation,
we compute the leading order in $n_l$ only.
In this approximation the operators $O_l$ and $O_s$ do not mix in
matching coefficients.
We need, in addition to the computation of the two-loop diagrams,
the heavy quark on-shell renormalization constant $Z_2$ up to NLO,
and the corresponding contributions to the anomalous dimension.

A single
bare operator $\mathrm{Q}$ is expanded over the basis of
renormalized operators
in the form
\begin{equation}
\mathrm{Q} = Z_QQ(\mu)+Z_{QE} E^{QCD}(\mu)
\end{equation}
therefore,
\begin{equation}
Q(\mu) =(\mathrm{Q}-Z_{QE} E^{QCD}(\mu))/Z_Q\, .
\end{equation}
Bare evanescent operators are expanded through
\begin{equation}
\mathrm{E^{QCD}}=Z_{EQ} Q(\mu)+Z_{EE} E^{QCD}(\mu)\, .
\end{equation}
Finally
$E^{QCD}(\mu)=0$ and the quantity $Z_{EQ}$ is designed so
that $\langle \mathrm{E^{QCD}}\rangle=Z_{EQ} \langle Q(\mu)\rangle$.
At LO the quantity
$Z_{EQ}$ has no poles in $\epsilon$ that makes the
renormalization nonminimal.

Arranging things so that $E^{\rm QCD}\to e^{\rm HQET}$
one need not consider NNLO
evanescent but NLO evanescent should be matched at one-loop order.
At matching, one has to make sure that the
one-loop matching of $E^{\rm QCD}$ is
safe and the tree-level of new evanescent is at least of
$O(\ep^2)$ in the physical sector.

We define the expansion of the coefficients as
\begin{eqnarray}
C_l &=& 1 + \frac{\alpha_s(m_b)}{4\pi} C_l^{(1)} + \left(\frac{\alpha_s(m_b)}{4\pi}\right)^2 C_l^{(2)}\,,\nonumber\\
C_s &=& \frac{\alpha_s(m_b)}{4\pi} C_s^{(1)} + \left(\frac{\alpha_s(m_b)}{4\pi}\right)^2 C_s^{(2)}\,.
\end{eqnarray}
At NNLO we
compute explicitly only the contribution of light fermion loops
that is gauge
invariant.

The leading order result goes at $\mu=m_b$~\cite{FHH:91,B:96,CFG:96}
\begin{equation}
C_l^{(1)} = \frac{-8N_c^2-9N_c+15}{2N_c}\,,\quad
C_s^{(1)} = -2(N_c+1)\,.
\end{equation}
At NNLO the result for the scalar operator is
\begin{equation}
C_s=-2(N_c+1) \frac{\alpha_s(m_b)}{4\pi} \left(1 -
  \frac{\alpha_s(m_b)}{4\pi} \frac{38}{9} T_F n_l\right)\, .
\end{equation}
Here $T_F=\frac{1}{2}$, $n_l$ is a number of light (massless) quarks.
Numerically the contribution gives a
rather reasonable shift of order 20\% for $n_l=4$.

The result for the vector operator $O_l$ reads
\begin{equation}
C_l^{(2)} = T_F n_l \left(1-\frac{1}{N_c}\right)
\left(5(N_c+3) \zeta_2 + \frac{11}{12}(17 N_c+25)\right)\,,
\end{equation}
and for $N_c=3$
\begin{eqnarray}
C_l^{(2)}=n_l\left(10 \zeta_2+\frac{209}{9}\right)\, ,
\quad \zeta_2=\frac{\pi^2}{6}\approx 1.64\ldots\, .
\end{eqnarray}

Note that we disagree with the entry $n_l$ in the two-loop anomalous
dimension of ref.~\cite{G:93}.
The matching coefficient which relates
the renormalized operators in QCD and HQET must be finite
at $\epsilon\to0$.
This requirement is satisfied when we use the anomalous dimension
of $O_l$ derived in Appendix~\ref{app:adim}.
It is not satisfied if we use the anomalous dimension from~\cite{G:93}.

The shift in the coefficient $C_l$
for $n_l=4$ is
\begin{eqnarray}
C_l= 1 + \frac{\alpha_s(m_b)}{4\pi}\left( -14+36\frac{\alpha_s(m_b)}{\pi}\right)
\end{eqnarray}
that amounts to 20\% of the NLO result
and
should be taken into account in precision analysis.
Thus, NNLO corrections can be large and shift the results of NLO
computation by at least
10--20\%. This is an argument that the NNLO
corrections require full computation.

\section{Operator product expansion in HQET for the sum rules analysis}
\label{Sect:SR}
In this section we discuss the role of evanescent
operators
in the sum rules (SR)
analysis for the matrix element of
the four-quark operator $O_l$
within HQET. This analysis is important for our computation of
bag parameters with three-loop
correlator in~\cite{Grozin:2016uqy}.
In the previous section we established that $E^{\rm QCD}$
matches to evanescent operators in HQET provided
the basis is chosen as in~(\ref{eq:HQET-good-basis})
which implies that only physical operators $O_l,O_s$ should be
considered in SR analysis.
However, for the choice of the basis as in~(\ref{eq:basis-HQET})
it is not the case.
Here we explicitly demonstrate that
the operator $E^{QCD}$ after matching to HQET gives no contribution
to SR if the basis~(\ref{eq:HQET-good-basis}) is used for
the computation
of the matching coefficients $C_l,C_s$.
Thus, this section gives an explicit demonstration of usefulness of
introducing correlated evanescent operators in QCD and HQET.
One has to consider the Buras evanescent operator  $E^{QCD}$
that is used for computing the Wilson
coefficient $C(\mu)$ at NLO in QCD.
It suffices to compute the LO contribution of the evanescent operator
to the three-point correlator.
It requires a two-loop calculation of the operator product expansion (OPE)
but the integral factorizes.

To evaluate the matrix element of the mixing we use
a vertex (three-point) correlation function~\cite{Chetyrkin:1985vj}.
This correlator reveals the factorizable structure of the matrix
element more clearly than the two-point function~\cite{Narison:1994zt}
but is significantly more difficult to compute at NLO
compared to the
calculation of the two-point function.
For the present analysis we however
set up a three-point sum rule in
HQET where the computational difficulties have been
solved~\cite{GL:09}.
We consider a correlator~\cite{Grozin:2016uqy}
\begin{equation}
K = \int d^d x_1\,d^d x_2\,e^{i p_1 x_1 - i p_2 x_2}
\langle 0 |T \tilde{\jmath}_2(x_2)
O_l(0) \tilde{\jmath}_1(x_1) | 0 \rangle
\label{SR:K}
\end{equation}
of the operator $O_l$.
Here we compute in dimensional regularization with
$d = 4 - 2 \varepsilon$ and with
anticommuting $\gamma_5$. The currents
\begin{equation}
\tilde{\jmath}_1 = \bar{h}_+ \gamma_5 d\, ,\quad
\tilde{\jmath}_2 = \bar{h}_- \gamma_5 d\,
\label{SR:j}
\end{equation}
interpolate the ground state of a $B$ meson in a static
approximation.

The bare QCD operator is
\begin{equation}
\mathrm{Q}=Z_Q(\alpha_s(\mu))Q(\mu)+Z_{QE} E^{QCD}(\mu)
\end{equation}
and $E$ is an evanescent operator
that contains $\Gamma^{[3]}$ and has
$t^a\otimes t^a$ color structure.
It suffices to match $E$ to HQET at LO.
Then one has to compute a correlator at LO as well that in our case
is given by a two-loop integral but of a factorized form
(just a product of two one-loop integrals).

Thus, one has to compute the
three-point correlator at LO and make sure
that there is no contribution of the evanescent operator.
Due to the color structure of $E$ only $\infty$-type topology or one-trace
part of the correlator survives.
It goes symbolically as
\begin{equation}
{\rm tr}(S_L(k)\Gamma H_+(k,\omega_1,v) \gamma_5S_L(l)\Gamma H_-(l,\omega_2,v)
\gamma_5)\, .
\end{equation}

Here $S_L(k)=\slashed{k}(1+\gamma_5)/k^2$ is a
light quark propagator,
$ H_\pm(l,\omega,v)=(1\pm \slashed{v})/(-2lv+\omega)$ is a heavy quark propagator,
$\Gamma$ is either $\Gamma^{[3]}$ or $\Gamma^{[1]}$ for the evanescent
operator. Note that one can also use the basis of
full gamma matrices for the evanescent
operators that results only in reshuffling the basis.

One finds for the contribution of a four-quark operator of a
general gamma structure
\begin{eqnarray}
&&K_{\Gamma^{[0]}}=2(kv)(lv)\{(-1)\}\mathcal{K}_0, \nonumber   \\
&&K_{\Gamma^{[1]}}=2(kv)(lv)\{(2-d)\}\mathcal{K}_0, \nonumber   \\
&&K_{\Gamma^{[2]}}=2(kv)(lv)\{(d-1)(d-2)\}\mathcal{K}_0, \nonumber   \\
&&K_{\Gamma^{[3]}}=2(kv)(lv)\{(d-1)(d-2)(d-6)\}\mathcal{K}_0\, ,
\end{eqnarray}
where $\mathcal{K}_0$ is a scalar structure of the integral
\begin{eqnarray}
\mathcal{K}_0=\frac{4}{(-2(lv)+\omega_1)(-2(kv)+\omega_2)l^2k^2}\, .
\end{eqnarray}
The $d$-depending factor of $\Gamma^{[3]}$ operator is
$-12+8\ep$.

We remind one that the evanescent operator is
\begin{equation}
E^{QCD}=\Gamma^{[3]}\otimes\Gamma^{[3]}
+(-6-2\ep)\Gamma^{[1]}\otimes\Gamma^{[1]}+O(\ep^2)
\end{equation}
and it matches onto HQET operators at LO
that retain its Dirac gamma structure and color structure.
Thus,
one finds the contribution of the  evanescent operator to the
correlator to be
\begin{eqnarray}
K_E&=&\mathrm{const} \rho(\omega_1)\rho(\omega_2)
\big(-12+8\ep+(-6-2\ep)(-2+2\ep)\big)
\nonumber   \\
&=&\mathrm{const} \rho(\omega_1)\rho(\omega_2)
\left(-12+8\ep+(12-8\ep)\right)=O(\ep^2)\, .
\end{eqnarray}
There is a pole in $\ep$ from the
renormalization constant and finite factors
$\rho(\omega_1)\rho(\omega_2)$
after taking
the imaginary parts in $\omega_{1,2}$.
All together this gives a total factor of
$\ep$ multiplying the finite result for the correlator
and the contribution of the  evanescent operator vanishes
as it should be.

Thus, the calculation in ref.~\cite{Grozin:2016uqy} is not affected by
the presence of evanescent operators as soon as they are defined
according to the standard rules used for the computation of the Wilson
coefficients. This fact has not been explicitly mentioned in
ref.~\cite{Grozin:2016uqy}.

In general, an HQET basis should respect the relation
\begin{eqnarray}
 \gamma_\perp^{[3]}\otimes \gamma_\perp^{[3]}
-3 \gamma_\perp^{[2]}\otimes \gamma_\perp^{[2]}+
(-6-2\ep)( \gamma_\perp^{[1]}\otimes \gamma_\perp^{[1]}
- \gamma_\perp^{[0]}\otimes \gamma_\perp^{[0]})=O(\ep^2)
\end{eqnarray}
for the
$t^a\otimes t^a$ color structure where only one trace
is possible for the three-point correlator.
Our choice of the basis agrees with this requirement.

\section{A comment on $B_s$ mixing}
\label{Sect:Flavor}
Mixing effects occur in the $B_d$ as well as in the $B_s$ system,
and in both systems the mixing parameters have been measured quite
precisely. In both cases, the mixing frequency is dominated by the top
quark and thus can be computed in terms of the local four-quark matrix
elements discussed above. Aside from the different Cabibbo--Kobayashi--Maskawa (CKM) factors
[$(V_{td} V_{tb}^*)^2$ compared to $(V_{ts} V_{tb}^*)^2$] the hadronic matrix
element involves an $s$ quark instead of a $d$ quark, and the states
need to be replaced accordingly. It has become customary to define
the matrix element
\begin{eqnarray} \label{ME}
\langle B_q |  (\bar q_L \gamma_\alpha b) \, (\bar q_L \gamma_\alpha
b) (\mu) |
{\bar B}_q \rangle
= \frac{2}{3} m_{B_q}^2  f_{B_q}^2 B_{B_q}
\end{eqnarray}
in terms of the corresponding decays constant $f_{B_q}$
multiplied by a bag factor  $B_{B_q}$
which is unity in naive factorization.
Phenomenological
CKM fits make use of lattice
calculations of the relevant hadronic matrix elements.
It turns out that
the ratio
\[
\xi^2=\frac{ f_{B_s}^2 B_{B_s}}{ f_{B_d}^2 B_{B_d}}
\]
can be computed quite precisely on the lattice, since many
systematic uncertainties cancel in the ratio.
The quantity $\xi$ in combination with
the perturbative calculation of
the Wilson coefficient is the basis for the
extraction of $V_{td} {V_{tb}}^*$ which
fixes one of the sides of the unitarity triangle.

In a recent paper \cite{Grozin:2016uqy} we have
discussed the matrix element~(\ref{ME})  for a $B_d$ meson,
using a QCD
sum rule within HQET. While in a lattice calculation
the decomposition of the matrix element into decay
constant and bag parameter is irrelevant, it is important
in the sum-rule calculation, since the contributing
Feynman diagrams can be uniquely attributed to either $f_{B_q}$ or
$B_{B_q}$.
Furthermore, the sum rule
allows us to estimate the deviation of $B_{B_q}$ from unity, which
eventually
leads to a quite precise result
for the bag factor $B_{B_d}$.

A similar estimate for the bag parameter $B_{B_s}$ requires one to take
into account $SU(3)$ breaking effects,
which are induced by the strange-quark mass $m_s$. In a sum rule, this
parameter appears, on the one hand, explicitly in
the perturbative calculation; on the other hand, it will also induce
the
difference between the strange- and the
light-quark condensates.

It is well known that the $SU(3)$ breaking is in general not small
as one can see for light mesons $K$ and $\pi$. For example,
the dependence on $m_s$
is well seen in the sum-rule calculation for the leptonic
decay constants $f_K$ and
$f_\pi$~\cite{Ovchinnikov:1985bs}.
It is not small as it emerges at the tree level and is of the
order $\frac{m_s}{\omega_c}\approx 0.15$ with
$m_s=150~\mathrm{MeV}$
and ${\omega_c}=1~\mathrm{GeV}$ is a typical scale of
sum-rules computation. Indeed,  experimentally we have
$(f_K-f_\pi)/f_\pi=0.19$, which is parametrically close to the
above estimate.

However, if we look at
bag parameters, we see that the leading term of the  sum-rules
computation is completely factorized and predicts $B=1$ for both
${B_s}$ and ${B_d}$ mesons.
This means in turn that the ratio of the two bag parameters
emerges only at NLO level and is
of the order
\begin{eqnarray}
\frac{\alpha_s}{\pi}\frac{m_s}{\omega_c}\approx 0.13\cdot 0.15=0.02\, .
\end{eqnarray}
Thus our best estimate is
\begin{eqnarray}\label{eq:bag-ratiox}
\frac{ B_{B_s}}{B_{B_d}}=1 \pm  0.02 \, .
\end{eqnarray}

This may be compared to the prediction from the lattice which can be derived
by using the lattice predictions for the decay constants
to be~\cite{Bazavov:2016nty}
\begin{eqnarray}
\frac{ B_{B_s}}{B_{B_d}}|_{latt}=1.033(31)(26)\,,
\end{eqnarray}
which is compatible with our observation.

We conclude that the deviation from unity of
the quantity $\xi$ is almost
completely driven by the $SU(3)$ in the leptonic decays
constants. There are
also recent sum-rule estimates for these matrix elements given
in~\cite{Gelhausen:2013wia}.  In particular,
we obtain for the ratio~\cite{Gelhausen:2013wia}
\begin{eqnarray}
\frac{f_{B_s}}{f_{B_d}}=1.17^{+0.03}_{-0.04}\, .
\end{eqnarray}

The contributions of power corrections
to the sum rules for  mixing analysis
are pretty
small~\cite{Ovchinnikov:1985bs,Pivovarov:2012zz,MPP:11,OP:88}.
Making use of~(\ref{eq:bag-ratiox})
we obtain
\begin{eqnarray}\label{eq:xi-sr}
\xi|_{\rm sr}=1.17^{+0.05}_{-0.06}\, ,
\end{eqnarray}
to be compared with the lattice result~\cite{Bazavov:2016nty}
\begin{eqnarray}\label{eq:xi-latt}
\xi|_{latt}=1.206(18)(6)\, .
\end{eqnarray}

We see that our result~(\ref{eq:xi-sr})
agrees with the lattice value~(\ref{eq:xi-latt})
within our uncertainties but it is less precise.
The main uncertainty comes from the ratio of decay constants where
the sum-rule estimate remains
less precise as the current lattice evaluations.

On the other hand, the QCD sum-rule estimate reveals the relative size of the
different contributions to the four-quark matrix elements
as well as to the parameter
$\xi$. The key observation is that naive factorization of the
four-quark matrix elements
is in fact a quite good approximation; QCD sum rules indicate
that the correction to this
assumption is small. This is
even more true for the ratio $\xi$, which is (up to a small
correction of the order of 2\%) driven by the $SU(3)$ breaking
in the leptonic decay
constants.
Indeed, by taking
the world average results for the decay
constants~\cite{Rosner:2015wva}
\begin{eqnarray}
\left(\frac{f_{B_s}}{f_{B_d}}\right)_{|_{\rm av}}=1.192(6)
\end{eqnarray}
we find
\begin{eqnarray}
\xi|_{mix}=\sqrt{\frac{ B_{B_s}}{B_{B_d}}}_{|_{\rm sr-est}}
\left(\frac{f_{B_s}}{f_{B_d}}\right)_{|_{\rm av}}=(1\pm 0.01)|_B
1.192(6)_{\rm av}=1.19(2)
\end{eqnarray}
which is comparable
with the lattice result~(\ref{eq:xi-latt}).

\section{Summary}
\label{Sect:Summary}
We discuss the problem of higher order corrections in HQET for
the analysis of mixing in sum rules.
We fix the basis of HQET operators with a set of evanescent
operators
necessary
in higher orders within computation in naive
dimensional regularization.
We have considered matching of QCD to HQET at NNLO where the
precise definition of evanescent operators is a must.
We have computed the contribution due to light fermion
loops and discovered that
our result disagrees with the entry of the anomalous dimension used
before.
We have computed leading $n_l$ contributions.
They happen to be large at the level of 10\%.
If there is no cancellation in the full result our numbers show that
the precision of matrix elements (ME) at the level of a few percent requires the matching
coefficients at NNLO. Note in passing that with the NNLO accuracy of
the leading term one may need to account also for nonleading terms of
HQE~\cite{KM:92}.

We have revisited the calculation of the
bag parameter in sum rules at three-loop level~\cite{Grozin:2016uqy}.
We stress that the form of the Hamiltonian in the physical sector
is not sufficient:
for any computation of ME (or Green functions) one needs to know
the set of related evanescent operators.
This makes the computation rather cumbersome as the
renormalized physical operators in such a scheme
have no meaning without an explicit form of the evanescent ones.
We show that our choice of evanescent operators
coordinated with the standard
ones in QCD does not affect our calculation of the three-loop
correction to the bag parameter in HQET~\cite{Grozin:2016uqy}.

We also discuss the flavor dependence of the bag parameters
for strange bottom meson.
Within the SR approach, in contrast to lattice, one can see the
anatomy of contributions and guarantee that the flavor shift is very
small.
This makes the prediction of the ratio $\xi$ dependable only on the
ratio of leptonic couplings that is rather precisely
known.

\section{Acknowledgment}
We thank U.Nierste for interest in the work and kind
indication on papers~\cite{Herrlich:1994kh,Herrlich:1996vf}.
A.G. is grateful to Siegen University for hospitality;
his work has been partially supported by the Russian Ministry of
Education and Science.
This work is supported by the DFG Research Unit FOR 1873
"Quark Flavour Physics and Effective Theories".

\begin{appendices}

\section{HQET bases for evanescent operators}
\label{app:basis}

The physical pair can be chosen differently than in the main text.
Let's take  $O_0, O^\prime_0$ that look rather symmetric
and differ by color arrangement only.
Then the reduction of operators up to rank three reads
\begin{eqnarray*}
O_1&=&-2O^\prime_0-O_0\,,\quad
O'_1=-O^\prime_0-2O_0\,,\\
O_2&=&2O_0+4O^\prime_0\,,\quad
O'_2=4O_0+2O^\prime_0\,,\\
O_3&=&-6O_0\,,\quad
O'_3=-6O^\prime_0\,,
\end{eqnarray*}
and this determines a new set of evanescent operators.

Yet another physical
basis is $O_+=O_0+O'_0$, $O_-=O_0-O'_0$.
These operators are Fierz eigenstates by construction with $\pm 1$
parity at tree level in four dimensions.
The reduction in this basis is
\[
O_1^+=-3O_+\,,\quad
O_1^-=O_-\,,\quad
O_2^+=6O_+\,,\quad
O_2^-=-2O_-\,,\quad
O_3^\pm=-6O_\pm\, .
\]
Note again that this is a pure four
dimensional reduction that leads to
evanescent operators of a minimal choice.

Clearly, the freedom of the definition of evanescents
is not only the choice of a physical pair but
deviation from minimality.
By adding a physical operator to an
evanescent with a coefficient vanishing in four-dimensional space
gives a
nonminimal basis. For example,
the shift of the evanescent $e$
with a physical operator $O$ with a coefficient of order $\ep$
\begin{eqnarray}
e\to e+(d-4)O
\end{eqnarray}
changes the basis and, therefore, anomalous dimensions and matching
coefficients for physical operators in higher orders.

Our working basis is
\begin{equation}
\label{eq:shift}
O_3\to O_3+2\ep(O_1-O_0)\,,\quad
O'_3\to O'_3+2\ep(O'_1-O'_0)\,.
\end{equation}
It means that one first does such a shift and then applies
the standard
minimal reduction.
Or, in addition to this,
\begin{eqnarray}
\label{eq:minbasisfin}
&&e_1=O_2+2O_1\,,\quad
e_2=O_3+6O_0\,,\quad
e_3=O'_0+\frac{1}{2}(O_0+O_1)\,,\nonumber\\
&&e_4=O'_1+\frac{3}{2}O_0-\frac{1}{2}O_1\,,\quad
e_5=O'_2-3O_0+O_1\,,\quad
e_6=O'_3-3 O_0-3O_1
\end{eqnarray}
is applied.

Combining two expansions
(\ref{eq:shift}) and (\ref{eq:minbasisfin})
one obtains the basis as
\begin{eqnarray}
O_2&=&-2O_1+e_1, \nonumber   \\
O_3&=&-6O_0+2\ep(O_1-O_0)+e_2+O(\ep^2), \nonumber   \\
O^\prime_0&=&-\frac{1}{2}(O_0+O_1)+e_3, \nonumber   \\
O^\prime_1&=&-\frac{3}{2} O_0+\frac{1}{2}O_1+e_4, \nonumber   \\
O^\prime_2&=&3 O_0-O_1+e_5, \nonumber   \\
O^\prime_3&=&3(O_0+O_1)+2\ep(O_1-O_0)+e_6+O(\ep^2)\,
\end{eqnarray}
and we have redefined $e_6$ compared to the minimal basis.
With such a substitution the QCD canonical evanescent operator
matches to pure
HQET evanescent with necessary order in $\ep$-expansion.

\section{Renormalization}
\label{app:ren}

Here we fix some notation.
The $\overline{\text{MS}}$ renormalized coupling constant is
\begin{equation}
\frac{g_0^2}{(4\pi)^{d/2}} = \mu^{2\epsilon} e^{\gamma\epsilon}
\frac{\alpha_s(\mu)}{4\pi} Z_\alpha(\alpha_s(\mu))\,,\quad
Z_\alpha = 1 - \beta_0 \frac{\alpha_s}{4\pi\epsilon} + \cdots\,,\quad
\beta_0 = \frac{11}{3} C_A - \frac{4}{3} T_F n_l\,.
\end{equation}
A renormalized operator $O(\mu)$ is related to the bare one $\mathrm{O}$
by $\mathrm{O} = Z(\alpha_s(\mu)) O(\mu)$.
Its anomalous dimension is
\begin{equation}
\gamma = \frac{d\log Z}{d\log\mu} = \gamma_0 \frac{\alpha_s}{4\pi}
+ \gamma_1 \left(\frac{\alpha_s}{4\pi}\right)^2 + \cdots\,,
\end{equation}
and the renormalization constant must have the form
\begin{equation}
\log Z = - \frac{1}{2} \gamma_0 \frac{\alpha_s}{4\pi\epsilon}
+ \frac{1}{4} \left(\beta_0 \gamma_0 - \gamma_1 \epsilon\right) \left(\frac{\alpha_s}{4\pi\epsilon}\right)^2
+ \cdots
\end{equation}
The operator $Q$~(\ref{eq:Q}) has
\begin{equation}
\gamma_0 = 6 \frac{N_c-1}{N_c}\,,\quad
\gamma_1 = - \frac{N_c-1}{2N_c} \left(\ldots - \frac{4}{3} n_l\right)\,.
\end{equation}

For the matching calculation we need the on-shell
renormalization constant $Z_2^{OS}$~\cite{Broadhurst:1991fy}:
\begin{eqnarray}
Z_2^{\text{os}} &=& 1
- C_F \frac{g_0^2 (m_b^{\text{os}})^{-2\epsilon} e^{-\gamma\epsilon}}{(4\pi)^{d/2} \epsilon}
\left[3 + 4 \epsilon + \left(\frac{\pi^2}{4} + 8\right) \epsilon^2 + \cdots\right]
\nonumber\\
&&{} + C_F T_F n_l \left(\frac{g_0^2 (m_b^{\text{os}})^{-2\epsilon} e^{-\gamma\epsilon}}{(4\pi)^{d/2} \epsilon}\right)^2
\left[2 + 9 \epsilon + \left(\frac{5}{3} \pi^2 + \frac{59}{2}\right) \epsilon^2 + \cdots\right]
+ \cdots\,,
\end{eqnarray}
where only the $n_l$ contribution to the $g_0^4$ term is written,
and $m_b^{\text{os}}$ is the mass in the on-shell scheme.

\section{The $n_l \alpha_s^2$ term\\
in the anomalous dimension of $O_l$}
\label{app:adim}

An easy way to calculate this anomalous dimension is to use infrared rearrangement.
We nullify all external momenta (including HQET residual ones)
and introduce a gluon mass as an IR regulator.
The gluon propagator with a light-quark loop insertion is transverse;
it is convenient to keep this property also for the propagator without insertions,
i.\,e., to use Landau gauge.
The free gluon propagator becomes
\begin{equation}
\frac{i}{m^2-k^2} \left(g_{\mu\nu} - \frac{k_\mu k_\nu}{k^2}\right)\,.
\label{adim:D0}
\end{equation}
The quark-loop insertion is $i \Pi(k^2) \left(k^2 g_{\mu\nu} - k_\mu k_\nu\right)$,
$\Pi(k^2) = \beta_0 A_0 e^{-\gamma\epsilon} D(\epsilon) (-k^2)^{-\epsilon}$,
where $A_0 = g_0^2/((4\pi)^{d/2} \epsilon)$
and $D(\epsilon) = 1 + \frac{5}{3} \epsilon + \cdots$
(we will keep only the $-\frac{4}{3} T_F n_l$ term in $\beta_0$).
Therefore, the gluon propagator with up to 1 quark-loop insertion is~(\ref{adim:D0}) times
\begin{equation}
Q = 1 + \Pi(k^2) \frac{-k^2}{m^2-k^2}\,.
\label{adim:Q}
\end{equation}

Let's first discuss the bilinear current $\tilde{\jmath}^\alpha = \bar{q} \gamma^\alpha h_+$.
Its matrix element is $(Z_h Z_q)^{1/2} V \gamma^\alpha$, where the vertex is
\begin{align}
V = 1 + C_F A_0 \epsilon \int \frac{d^d k}{i\pi^{d/2}}
\frac{\rlap/v \rlap/k - k\cdot v}{k\cdot v\,(m^2-k^2) (-k^2)} Q\,.
\label{adim:j}
\end{align}
The vector integral with $k^\mu$ in the numerator can be directed only along $v^\mu$,
and we may substitute $k^\mu \to k\cdot v\,v^\mu$.
Loop corrections vanish, and we get
\begin{equation}
\tilde{\gamma}_j = \frac{1}{2} \left(\gamma_h + \gamma_q\right)\,,
\label{adim:gammaj}
\end{equation}
where $\gamma_{h,q}$ are in Landau gauge.
In this way we easily reproduce
the $\alpha_s$ and $n_l \alpha_s^2$ terms
in $\tilde{\gamma}_j$~\cite{BG:91},
\begin{equation}
\tilde{\gamma}_j=-3C_F\frac{\alpha_s}{4\pi}
+n_lC_F\frac{5}{3}\left(\frac{\alpha_s}{4\pi}\right)^2\,.
\end{equation}
Now we turn to the operator $O_l$.
Its matrix element is $Z_h Z_q V$, where the vertex is
\begin{equation}
V = T_1 O
+ 2 \raisebox{-10.7mm}{\includegraphics{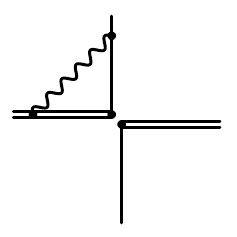}}
+ 2 \raisebox{-10.7mm}{\includegraphics{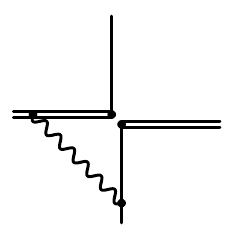}}
+ \raisebox{-10.7mm}{\includegraphics{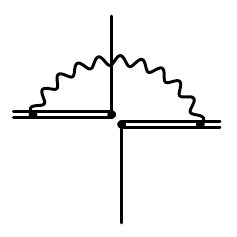}}
+ \raisebox{-10.7mm}{\includegraphics{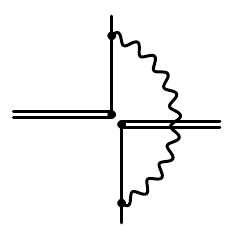}}\,,
\label{adim:V}
\end{equation}
where $O = \gamma^\alpha \otimes \gamma_\alpha$.
The color structure of the tree diagram is $T_1 = \delta_{a_1}^{b_1} \delta_{a_2}^{b_2}$
where $a_{1,2}$ are the color indices of the heavy external legs,
$b_{1,2}$ are those of the light legs,
and 1, 2 number the fermion lines.
The first loop diagram in~(\ref{adim:V}) is the same as the one for the current $\tilde{\jmath}$,
and hence it vanishes.
The second diagram differs from the first one only by the color factor:
instead of $C_F T_1$, it is now
\begin{equation}
T = T_F \left(T_2 - \frac{T_1}{N_c}\right)\,,\quad
T_2 =  \delta_{a_1}^{b_2} \delta_{a_2}^{b_1}\,;
\label{adim:T}
\end{equation}
it also vanishes.

The heavy--heavy diagram in~(\ref{adim:V}) is
\begin{equation*}
T O A_0 \epsilon \int \frac{d^d k}{i\pi^{d/2}} \frac{Q}{m^2-k^2}
\left[\frac{1}{(k\cdot v)^2} + \frac{1}{-k^2}\right]\,.
\end{equation*}
Averaging over $k$ directions, we may substitute $1/(k\cdot v)^2 \to -(d-2)/k^2$~\cite{BG:95}
(this has been rigorously proved in~\cite{GSS:06}).
This gives
\begin{equation*}
T O A_0 \epsilon (d-1) \int \frac{d^d k}{i\pi^{d/2}} \frac{Q}{(m^2-k^2) (-k^2)}
= T O A_0 m^{-2\epsilon} \Gamma(1+\epsilon) \frac{3-2\epsilon}{1-\epsilon}
\left[1 + \frac{1}{2} \beta_0 A_0 m^{-2\epsilon} e^{-\gamma\epsilon} D'(\epsilon)\right]\,,
\end{equation*}
where
\begin{equation*}
D'(\epsilon) = D(\epsilon) \frac{1+2\epsilon}{\cos(\pi\epsilon)}\,.
\end{equation*}
Re-expressing this result via the renormalized
\begin{equation*}
A = \frac{\alpha_s(m)}{4\pi\epsilon}\,,
\end{equation*}
we finally obtain
\begin{equation}
T O A \frac{e^{\gamma\epsilon} \Gamma(1+\epsilon)}{1-\epsilon} (3-2\epsilon)
\left[1 - \frac{1}{2} \beta_0 A \left(2 - D'(\epsilon)\right)\right]\,.
\label{adim:hh}
\end{equation}

The light--light diagram in~(\ref{adim:V}) is
\begin{equation*}
T A_0 \epsilon \int \frac{d^d k}{i\pi^{d/2}} \frac{Q}{(m^2-k^2) (-k^2)}
\left[\frac{\gamma^\mu \rlap/k \gamma^\alpha \otimes \gamma_\mu \rlap/k \gamma_\alpha}{-k^2} + O\right]\,.
\end{equation*}
We may average over $k$ directions: $k^\mu k^\nu \to (k^2/d) g^{\mu\nu}$, and obtain
\begin{equation*}
T \left(O - \frac{O_3}{d}\right) A_0 \epsilon
\int \frac{d^d k}{i\pi^{d/2}} \frac{Q}{(m^2-k^2) (-k^2)}\,,
\end{equation*}
where $O_3 = \gamma^\mu \gamma^\nu \gamma^\alpha \otimes \gamma_\mu \gamma_\nu \gamma_\alpha$.
Neglecting the evanescent operator
(which does not contribute to the $n_l \alpha_s^2$ term in the anomalous dimension)
we may replace $O_3 \to (16-4\epsilon) O$ and get
\begin{equation*}
 - T O A_0 m^{-2\epsilon} \Gamma(1+\epsilon) \frac{6-\epsilon}{(2-\epsilon) (1-\epsilon)}
\left[1 + \frac{1}{2} \beta_0 A_0 m^{-2\epsilon} e^{-\gamma\epsilon} D'(\epsilon)\right]\,,
\end{equation*}
or finally
\begin{equation}
- T O A \frac{e^{\gamma\epsilon} \Gamma(1+\epsilon)}{1-\epsilon} \frac{6-\epsilon}{2-\epsilon}
\left[1 - \frac{1}{2} \beta_0 A \left(2 - D'(\epsilon)\right)\right]\,.
\label{adim:ll}
\end{equation}

The renormalized operators $O_l' = O_l$,
so that we may replace $T \to T_F (1-1/N_c) T_1$.
Combining~(\ref{adim:hh}) with~(\ref{adim:ll})
we see that the vertex~(\ref{adim:V})
is $T_1 O V$ with
\begin{equation}
V = 1 - 2 T A \epsilon \frac{e^{\gamma\epsilon} \Gamma(1+\epsilon)}{1-\epsilon}
\frac{3-\epsilon}{2-\epsilon}
\left[1 - \frac{1}{2} \beta_0 A \left(2 - D'(\epsilon)\right)\right]\,.
\label{adim:Vres}
\end{equation}
We have $\log V = \log Z_V + \log V_r$,
where negative powers of $\epsilon$ go to $\log Z_V$,
while non-negative ones to $\log V_r$.
Taking logarithm of $V$~(\ref{adim:Vres}) we note that the square of the 1-loop term
does not contribute to the $n_l \alpha_s^2$ structure,
and $\log Z_V = \frac{3}{2} T \beta_0 A^2 \epsilon$
(we keep only the $n_l$ term in $\beta_0$).
The anomalous dimension of the operator $O_l$ is
\begin{equation*}
\tilde{\gamma} = 2 \tilde{\gamma}_j + \Delta\gamma\,,\quad
\Delta\gamma = \frac{d\log Z_V}{d\log\mu} = - 6 T \beta_0 \left(\frac{\alpha_s}{4\pi}\right)^2\,.
\end{equation*}
Finally, we arrive at
\begin{equation}
\Delta\gamma = 8 \left(1 - \frac{1}{N_c}\right) T_F^2 n_l \left(\frac{\alpha_s}{4\pi}\right)^2\,.
\label{adim:Dg}
\end{equation}

The result of~\cite{G:93} (where $T_F=\frac{1}{2}$) leads to
\begin{equation}
\Delta\gamma|_G = 2 \left(1 - \frac{1}{N_c}\right) \left(\frac{N_c}{3} + 1\right) n_l
\left(\frac{\alpha_s}{4\pi}\right)^2\,.
\label{adim:Gimenez}
\end{equation}
Our result~(\ref{adim:Dg}) differs from it by the absence of
the term $N_c/3$ in the bracket.
This color structure cannot appear in the
diagrams which contribute to the $n_l \alpha_s^2$ term
in $\Delta\gamma$.
Probably, other structures, without $n_l$,
which appear in the 2-loop $\Delta\gamma$,
also need rechecking.

\end{appendices}

\end{document}